\documentclass[11pt,a4paper]{article}
\usepackage{jheppub_kim}

\usepackage{pdflscape}
\usepackage{amsmath}
\usepackage{amssymb}
\usepackage{dcolumn}
\usepackage{bm}
\usepackage{color}
\usepackage{epsfig}
\usepackage{amsfonts}
\usepackage{graphicx}
\usepackage{subfigure}
\usepackage{dcolumn}

\newcommand{\be}{\begin{equation}}
\newcommand{\ee}{\end{equation}}
\newcommand{\bea}{\begin{eqnarray}}
\newcommand{\eea}{\end{eqnarray}}

\def\be{\begin{equation}}
\def\ee{\end{equation}}
\def\bea{\begin{eqnarray}}
\def\eea{\end{eqnarray}}

\begin{document}

\title{Effects of Hyperscaling Violation and Dynamical exponents on Drag Force }
\author[a]{M. Kioumarsipour}
\author[a]{J. Sadeghi}

\affiliation[a]{Sciences Faculty, Department of Physics, University of Mazandaran, Babolsar, Iran}

\emailAdd{m.kioumarsipour@stu.umz.ac.ir}
\emailAdd{pouriya@ipm.ir}

\abstract{\\
In this paper, we calculate drag force of a charged particle in systems with a hyperscaling violation, by using the AdS/CFT correspondence. We obtain energy loss, friction coefficient, diffusion and quasi normal modes of a heavy point particle. Here, we consider two cases. The first, we assume the heavy point particle moves with constant velocity $ v $. In the second case, we assume the heavy point particle rotates along a circle of radius $ L $ with angular velocity $ \omega $. Also, we add some electromagnetic field and obtain the energy loss. Finally, we draw some figures for the drag force with respect to the velocity and the temperature. As well, we show how the drag force and energy loss change with the hyperscaling violation parameter $ \theta $, dynamical parameter $ z $, mass $ m $ and charge $ Q $.

\textbf{Keywords:} AdS/CFT correspondence, Quark-Gluon plasma, Hyperscaling violation, Drag Force.}

\maketitle

\section{Introduction}
The AdS/CFT correspondence \cite{maldesena,witten,gubser} is a powerful and practical method to understand many insight into the dynamics of strongly-coupled gauge theories. Such correspondence gives us duality between supergravity theories that live in $ d+1 $-dimensional anti de Sitter (AdS) spacetimes with quantum field theories that live on the boundary $ d $-dimension. If we consider AdS geometry, the dual field theory will be conformal invariant and a strongly coupled theory with a UV fixed point. Recently, some systems are observed whose dual field theory is not conformally invariant, but they are still scale invariant. These cases correspond to geometries that are not asymptotically AdS. Such theories are of great importance in condensed matter physics \cite{condense}. The most important example is Lifshitz fixed point which has anisotropic scale symmetry.
It means that such systems are spatially isotropic but scaled anisotropically in the time direction by a dynamical exponent $ z $. So, such system will be invariante under the following scale transformations,
\begin{equation}
(t,\vec{x})\rightarrow (\lambda^{z}t,\lambda\vec{x}).
\end{equation}
In the case of $ z=1 $, theory has a relativistic scale invariance, and in the case $ z\neq 1 $, theory is non-relativistic because it is scaled differently in time and space. In fact, in order to have such theories, the Einstein gravity must be coupled to the abelian gauge ﬁelds \cite{lif1,lif2}. If gravity theory includes a scalar field as well, the full class of metrics which exhibit scaling properties will emerge \cite{lif3,lif4,lif5,lif6,lif7,lif8,lif9}. Including both these fields, one may have a metric as following form \cite{alii},
\begin{equation}
ds^{2}=r^{-2\theta /d}\left( -r^{2z}dt^{2}+\dfrac{dr^{2}}{r^{2}}+r^{2}d{\bf x}^{2}\right) ,\label{tt} 
\end{equation}
The (\ref{tt}) is spatially homogeneous and not invariant under scaling  but exhibits the following scale transformations,
\begin{equation}
t\rightarrow \lambda^{z}t,~~~~r\rightarrow \lambda^{-1}r,~~~~x_{i}\rightarrow \lambda x_{i},~~~~ds_{d+2}\rightarrow \lambda^{\theta/2}ds_{d+2}.
\end{equation}
An interesting property of such metrics is that they break conformal invariance. Already, for breaking conformal invariance was used of hard-wall and soft-wall approximations that suffered of some problems. For studying holography aspects of hyperscaling violation metric background one can see \cite{null}.

One of the most important and interesting problems in investigating quark gluon plasma (QGP) produced at RHIC and LHC is the study of energy loss and drag force. For this purpose one should consider the motion of a heavy probe in the thermal medium which corresponds to a hanging string from the boundary to the horizon. First study has been done in \cite{etl,herzog} that an external quark moving through $ \mathcal{N} = 4 $ supersymmetric Yang-Mills plasma is considered. In \cite{dr1,dr2,dr3} authors have considered a moving quark in the thermal plasma at the $ \mathcal{N}=2$ supergravity theory. The drag force and diffusion coefficient of a moving heavy quark in a D-instanton background have been studied in \cite{zha}. The D-instanton density increases the drag force, decreases the diffusion coefficient and makes the medium less viscous. The authors in \cite{shan} has considered the non-relativistic Brownian motion of a heavy quark in a strongly coupled, thermal, anisotropic Yang-Mills plasma and analyzed the effects of anisotropy on the drag coefficient and the diffusion constant for both the directions. The effect of finite-coupling corrections \cite{mada} and the effects of charge and finite 't Hooft coupling correction \cite{fada} on the drag force have been studied and the diffusion coefficient of heavy quarks is calculated from the drag force.
 In \cite{field} is added a constant electromagnetic ﬁeld on the D-brane and calculated drag force in a thermal plasma of $ \mathcal{N}=4 $ super Yang-Mills theory.
The experiments indicate that a strong electromagnetic field can affect on the QGP produced in heavy-ion collisions. Also, it is interesting to study the effect of the magnetic field on various quantities holographically \cite{hol1,hol2,hol3,hol4}. In \cite{hol4} authors studied the effect of a constant magnetic field $ \mathcal{B} $ on the drag force. They resulted that the energy loss of a massless quark moving or a heavy quark moving, in the presence of strong magnetic field $ \mathcal{B} $, is linearly dependent on $ \mathcal{B} $. Also, in \cite{mag} authors investigated the effect of a constant magnetic field $ \mathcal{B} $ on a heavy quark moving through a strongly-coupled $ \mathcal{N}=4 $ supersymmetric Yang-Mills plasma that the quark can move transverse and parallel to $ \mathcal{B} $. They concluded that the drag force is linearly dependent on $ \mathcal{B} $ for trensverse case and increases monotonously with increasing $ \mathcal{B} $ for parallel case that these results are in contradiction with \cite{field}, interestingly.
For more studies in different backgrounds one can see \cite{E,R}. The extremal black holes and the energy loss of a moving heavy point particle at zero temperature and ﬁnite charge density is studied in \cite{bi}. Also, the drag force in asymptotically Lifshitz spacetimes and in non-relativistic Lifshitz or Schrodinger theories has been studied in \cite{z2,z1,z3,z4,z5}. An interesting property of non-relativistic theories is that in these theories the drag force is not zero even at zero temperature. Theories with hyperscaling violation has been reported in \cite{hyp1,hyp2,hyp3,ata}. Although in these theories temperature is zero, the drag force is non-zero. Immediately after the collision the created system is far from equilibrium and plasma is anisotropic. In \cite{aniso,aniso1} studied the drag force on a heavy quark moving through a strongly coupled, anisotropic $ \mathcal{N}=4 $ plasma. In \cite{fieldd} is calculated the drag force acting on a moving quark through an anisotropic $ \mathcal{N}=4 $, $ SU(N) $ super Yang-Mills plasma in the presence of a $ U(1) $ chemical potential. Another interesting challenge is a rotating heavy quark along a circle of radius $ L $ with a constant angular frequency $ \omega $. It is dual with a string that coils in the bulk. The energy loss of such rotating particle has been studied in \cite{rot,rott,rot2}. The energy loss of a rotating quark at constant velocity, through an anisotropic strongly coupled $ \mathcal{N}=4 $ plasma has been studied in \cite{rot1}.

The most important purpose of this paper is that we employ all the parameters of theory and show how they affect our results. We consider the most interesting background that has mass, charge, dynamical exponent $ z $ and hyperscaling violation exponent $ \theta $. Here, without using the hard and soft-wall method we utilize the hyperscaling violation metric background. Then, we study the dynamics of a heavy quark which moves with constant velocity and rotates with constant angular frequency in the non-relativistic background with various values of $ m $, $ Q $, $ z $ and $ \theta $. All the above information makes us motivated to study the drag force and energy loss of this case. This paper is organized as follows. In section 2, we investigate the drag force on a moving quark and analyze the effect of mass $ m $, charge $ Q $, dynamical exponent $ z $ and hyperscaling violation $ \theta $ on it. In section 3, we calculate the quasi normal modes of string. After that, in section 4 we add a constant electromagnetic ﬁeld to the system and study its effect on the drag force. In section 5, we consider a circular motion of the quark and obtain its energy loss. Finally, in section 6 we summarize our results. 

\section{EMD Solutions and Drag Force}
In this section, we are going to study the drag force of a string that hangs from the boundary to the horizon in a hyperscaling violating metric background. From, the AdS/CFT correspondence, we know that the endpoint of this string corresponds to a heavy moving point particle. It is known that the velocity of the particle decreases due to viscosity of the medium, therefore, it feels a drag force. The drag force is associated with the friction coefficient $ \mu $, given by the Langevin equation, 
\begin{equation}
\dfrac{dp}{dt}=-\mu p+F
\end{equation}
subjected to a diriving force $ F $. When the quark moves with a constant velocity, then $ dp/dt=0 $ and the drag force will be equal to the driving force. 

Now, we consider the corresponding black hole solution of the Einstein-Maxwell-Dilaton theory. The solution of the Einstein-Maxwell-Dilaton system with hyperscaling violation is introduced by \cite{alii},
\begin{equation}
ds^{2}=r^{2\alpha}\left( -r^{2z}f(r)dt^{2}+\dfrac{dr^{2}}{r^{2}f(r)}+r^{2}d{\bf x}^{2}\right), ~~~~\alpha:=-\theta /d, \label{1}
\end{equation}
where $ z $ is dynamical exponent and $ \theta $ is hyperscaling violation exponent. The charged black hole solution is,
\begin{equation}
f(r)=1-\dfrac{m}{r^{z+d-\theta}}+\dfrac{Q^{2}}{r^{2(z+d-\theta-1)}},
\end{equation}
where $ m $ and $ Q $ are related to the mass and charge of the black hole respectively. Radius of horizon, $ r_{H} $, can be obtained by setting $ f=0 $ as,
\begin{equation}
r_{H}^{2(d+z-\theta -1)}-mr_{H}^{d+z-\theta -2}+Q^{2}=0.
\end{equation}
The Hawking temperature $ T $ is given by,
\begin{equation}
T=\dfrac{r_{H}^{z}(d+z-\theta)}{4\pi}\left( 1-\dfrac{(d+z-\theta -2)Q^{2}}{d+z-\theta}r_{H}^{2(\theta -d-z+1)}\right).\label{TT} 
\end{equation}

As we know for having a physically sensible dual field theory, the null energy condition (NEC) should be satisfied at least from the gravity side. So, in that case for an arbitrary null vector $ N^{\mu} $ one can write $ T_{\mu\nu}N^{\mu}N^{\nu}\geq 0 $  \cite{null,null1}. Applying the NEC on (\ref{1}) leads to following condition,
\begin{equation}
(d-\theta)(d(z-1)-\theta)\geq 0.\label{a} 
\end{equation}
This condition yields the allowed values for $ (z,\theta) $ so that the given gravity dual will be consistent. The Lorentz invariant theory corresponds to $ z=1 $ and the (\ref{a}) implies $ \theta \leq 0 $ or $ \theta\geq d $. The range of $ \theta > d $ is just allowed based on the NEC, in this range there is some instabilities in the gravity side \cite{null}. Therefore, we only consider the range of $ \theta\leq d $. For $ \theta =0 $, that demonstrates a scale invariant theory, we reachieve the known result $ z\geq 1 $.

If we consider case $ \theta\leq d $, then we conclude the following dynamical exponent from the condition (\ref{a}),
\begin{equation}
z\geq 1+\frac{\theta}{d}. \label{b} 
\end{equation}
If $ \theta =d, d-1 $ and $ d/2 $, therefore $ z\geq 2, 2-1/d $ and $ 3/2 $ respectively. \\
As shown in \cite{null}, the range of $ d-1\leq\theta\leq d $ displays novel phases that reveal new violations of the area law that interpolate between logarithmic and linear behaviors. The case $ \theta=d $ points out an extensive violation of the entanglement entropy. In theories with $ \theta=d-1 $ the holographic entanglement entropy exhibits a logarithmic violation of the area law which indicates the model must have a Fermi surface \cite{fermi,fermi1}.

Now, it turns to introduce the Nambu-Goto action for the investigation of the energy loss and other physical phenomena.
So, as we know the dynamics of an open string is governed by the Nambu-Goto action,
\begin{eqnarray}
S&=&-T_{0}\int d\tau d\sigma\mathcal{L}\nonumber\\
&=&-T_{0}\int d\tau d\sigma\sqrt{-g},
\end{eqnarray}
where $ T_{0}=\frac{1}{2\pi \alpha^{\prime}} $ is the string tension and the string worldsheet coordinates are shown by $ \tau $ and $ \sigma $. Determinant of the worldsheet metric, $ g_{ab} $, is represented by $ g $ which is induced by space-time metric $ G_{\mu\nu} $, (\ref{1}).
We restrict the motion of the string only in $ x $ direction and use the static gauge, namely $ t=\tau $ and $ \sigma=r $. With these assumptions $ x(t,r) $ describes the world-sheet of the string. Therefore, one can write,
\begin{equation}
-g=r^{4\alpha}\left( r^{2z-2}+r^{2z+2}f(r)x^{\prime2} -\dfrac{\dot x^{2}}{f(r)}\right) ,
\end{equation}
and the equation of motion is obtained by following equation,
\begin{equation}
\dfrac{\partial}{\partial r}\left( \dfrac{r^{\beta}f(r)x'}{\sqrt{-g}}  \right) +\dfrac{\partial}{\partial t}\left( \dfrac{-r^{4\alpha}f(r)^{-1}\dot{x}}{\sqrt{-g}}  \right)=0,~~~~ \beta:=4\alpha +2z+2 .\label{2} 
\end{equation}
Now, we should calculate the following canonical momentum densities,
\begin{equation}
 \left(
\begin{array}{ccc}
\pi^{0}_{x} & \pi^{1}_{x} & \\
\pi^{0}_{r} & \pi^{1}_{r} & \\
\pi^{0}_{t} & \pi^{1}_{t} &
\end{array} \right) =\dfrac{T_{0}r^{4\alpha}}{\sqrt{-g}} \left(
\begin{array}{ccc}
\dfrac{\dot{x}}{f(r)} & -r^{2z+2}f(r)x' &  \\
-\dfrac{\dot{x}x' }{f(r)} & -r^{2z-2}+\dfrac{\dot{x}^{2} }{f(r)} &  \\
-r^{2z+2}(r^{-4}+f(r)x'^{2}) &r^{2z+2}f(r)\dot{x}x'  &
\end{array} \right). \label{3} 
\end{equation}
Using above relations, we can obtain the drag force, the energy loss and also the total energy and momentum of the string which  are described by following relations, 
\begin{eqnarray}
E&=&-\int_{r_{H}}^{r_{m}} dr~\! \pi_{t}^{0},\nonumber\\
p&=&\int_{r_{H}}^{r_{m}} dr~\! \pi_{x}^{0}. \label{4} 
\end{eqnarray}

The moving quark in thermal plasma is dual with a string which hangs from the $ D $-brane to the horizon of the black hole.
In order to find the drag force, we can investigate three cases for the equation of motion (\ref{2}). In the first case which is the simplest case, we consider a static quark in the thermal plasma which is dual with a string stretched straightforwardly from $ r_{m} $ on the $ D $-brane to the black hole horizon at $ r=r_{H} $, $ x(t,r)=x_{0} $ where $ x_{0} $ is constant. According to the relations (\ref{3}) and (\ref{4}), the total momentum and the drag force vanish and the total energy of the string can be obtained by following equation,
\begin{equation}
E=-\int_{r_{H}}^{r_{m}} \pi_{t}^{0} dr =T_{0}\int_{r_{H}}^{r_{m}} r^{2\alpha +z-1} dr=\dfrac{T_{0}}{2\alpha +z}\left( r_{m}^{2\alpha +z}-r_{H}^{2\alpha +z} \right).\label{EE} 
\end{equation}
In range of $ \theta\leq d $ from (\ref{b}) we have $ 1+\theta/d\leq z $ and hence by considering $ z=1+\theta/d $ we obtain,
\begin{equation}
E=\dfrac{T_{0}d}{d-\theta}\left( r_{m}^{1-\theta /d}-r_{H}^{1-\theta /d}\right).
\end{equation}
As we see, the total energy depends on the dimension of space and hyperscaling violation factor explicity. Now we consider different cases:
\begin{itemize}
\item $ \mathbf{\theta=d/2} $\\
In the special case $ \theta=d/2 $, from (\ref{EE}) we obtain,
\begin{equation}
E=\dfrac{T_{0}}{z-1}\left( r_{m}^{z-1}-r_{H}^{z-1}\right).
\end{equation}
In this case the energy depends on dynamical exponent explicity. For the case $ z=2 $, we find $ E=T_{0}\left( r_{m}-r_{H}\right) $, which equals to energy of the string of $ \mathcal{N}=4 $ supersymmetric Yang-Mills \cite{herzog}.
\item $ \mathbf{\theta=d-1} $\\
In this case from (\ref{EE}) we have,
\begin{equation}
E=\dfrac{T_{0}}{z-2+\frac{2}{d}}\left( r_{m}^{z-2+\frac{2}{d}}-r_{H}^{z-2+\frac{2}{d}}\right),
\end{equation}
Here, the energy is dependent on dimension of space and dynamical exponent. If we consider $ z=2 $, we can rewrite,
\begin{equation}
E=\dfrac{T_{0}d}{2}\left( r_{m}^{2/d}-r_{H}^{2/d}\right),
\end{equation}
Phrase (\ref{b}) indicates $ z\geq 2-1/d $. By considering $ z= 2-1/d $ we find,
\begin{equation}
E=T_{0}d \left( r_{m}^{1/d}-r_{H}^{1/d}\right). 
\end{equation}

\item $ \mathbf{\theta=d} $\\
If $ \theta=d $, we achieve,
\begin{equation}
E=\dfrac{T_{0}}{z-2}\left( r_{m}^{z-2}-r_{H}^{z-2}\right),
\end{equation}
From (\ref{b}) we find $ z\geq2 $, considering $ z=3 $ we obtain,
\begin{equation}
E=T_{0}\left( r_{m}-r_{H}\right),\label{mnm} 
\end{equation}
this case also equals to energy of string of $ \mathcal{N}=4 $ supersymmetric Yang-Mills \cite{herzog}.
\end{itemize} 
If we set $ 2\alpha +z $ in (\ref{EE}) equal to $ 1 $, we can obtain a relation between $ \alpha $ and $ z $,
\begin{equation}
\alpha =\dfrac{1-z}{2},
\end{equation}
since $ \alpha =-\frac{\theta}{d} $, we have,
\begin{equation}
\theta =\dfrac{d}{2}(z-1).\label{k} 
\end{equation}
Also, in this case, the energy of string reduces to the energy of string of $ \mathcal{N}=4 $ supersymmetric Yang-Mills. In (\ref{k}) if we set $ \theta=d $, we regain $ z=3 $, this case is the same as (\ref{mnm}).\\
The energy (\ref{EE}) will be equal the rest mass of the quark in the zero temperature limit. According to the (\ref{TT}), that $ r_{H} $ is proportional to the temperature, we find that,
\begin{equation}
E=M_{rest}=\dfrac{T_{0}}{2\alpha +z}(r_{m}^{2\alpha +z}).
\end{equation}
 
The second case that we consider is the straight string which moves with constant velocity, $ x(t,r)=x_{0}+vt $. In this case, while the energy and momentum are nonzero on the string world sheet, the square root quantity is negative, therefore we have imaginary the action, energy and momentum, and so the motion is not physical.

The third case, we consider $ x(t,r)=x(r)+vt $ which is the most physical time-dependent solution for the equation of motion (\ref{2}) and describes a curved string that moves with a constant velocity $ v $. In this case, the equation of motion (\ref{2}) reduces as following,
\begin{equation}
\dfrac{\partial}{\partial r}\left( \dfrac{r^{\beta}f(r)x'}{\sqrt{-g}} \right) =0,
\end{equation}
and we find,
\begin{eqnarray}
&&Cv=\dfrac{r^{\beta}fx'}{\sqrt{-g}},\nonumber\\
&&-g=\dfrac{r^{2\beta- 4}f-r^{\beta+ 4\alpha}v^{2}}{r^{\beta}f-C^{2}v^{2}},\nonumber\\
&&x'^{2}=\dfrac{C^{2}v^{2}(r^{\beta -4}f-r^{4\alpha}v^{2)}}{r^{\beta}f^{2}(r^{\beta}f-C^{2}v^{2})},\label{xx} 
\end{eqnarray}
where $ C $ is the constant of motion. In order to have the physical action, energy and momentum the constant $ C $ must satisfy the reality condition $ -g $, which is given by,
\begin{equation}
C=\pm r_{c}^{2\alpha +2},
\end{equation}
where $ r_{c} $ is the minimum critical radius which is root of the following equation,
\begin{equation}
r_{c}^{2(z-1)}f_{c}-v^{2}=0,\label{c} 
\end{equation}
where $ f_{c} $ is the function $ f(r) $ in $ r_{c} $. This equation is complex and should solve numerically. Now, we can obtain the energy and momentum current of the string from the $ D $-brane to the horizon,
\begin{equation}
\pi_{x}^{1}=-T_{0}Cv,~~~~~~~\pi_{t}^{1}=T_{0}Cv^{2}. \qquad
\end{equation}
As we see $ \pi_{t}^{1} $, loss of the energy current of quark by string, is proportional to $ C $ and we should determine the sign. Since the quark moves in the thermal plasma and pulls the string, the energy and momentum current flow from the quark to the horizon along string, therefore the sign of $ C $ must be positive. The negative sign of $ C $ shows a case in which the string moves and trails the quark. This case is not a physical situation.\\
\begin{figure}
\begin{center}$
\begin{array}{cc}
\includegraphics[width=74 mm]{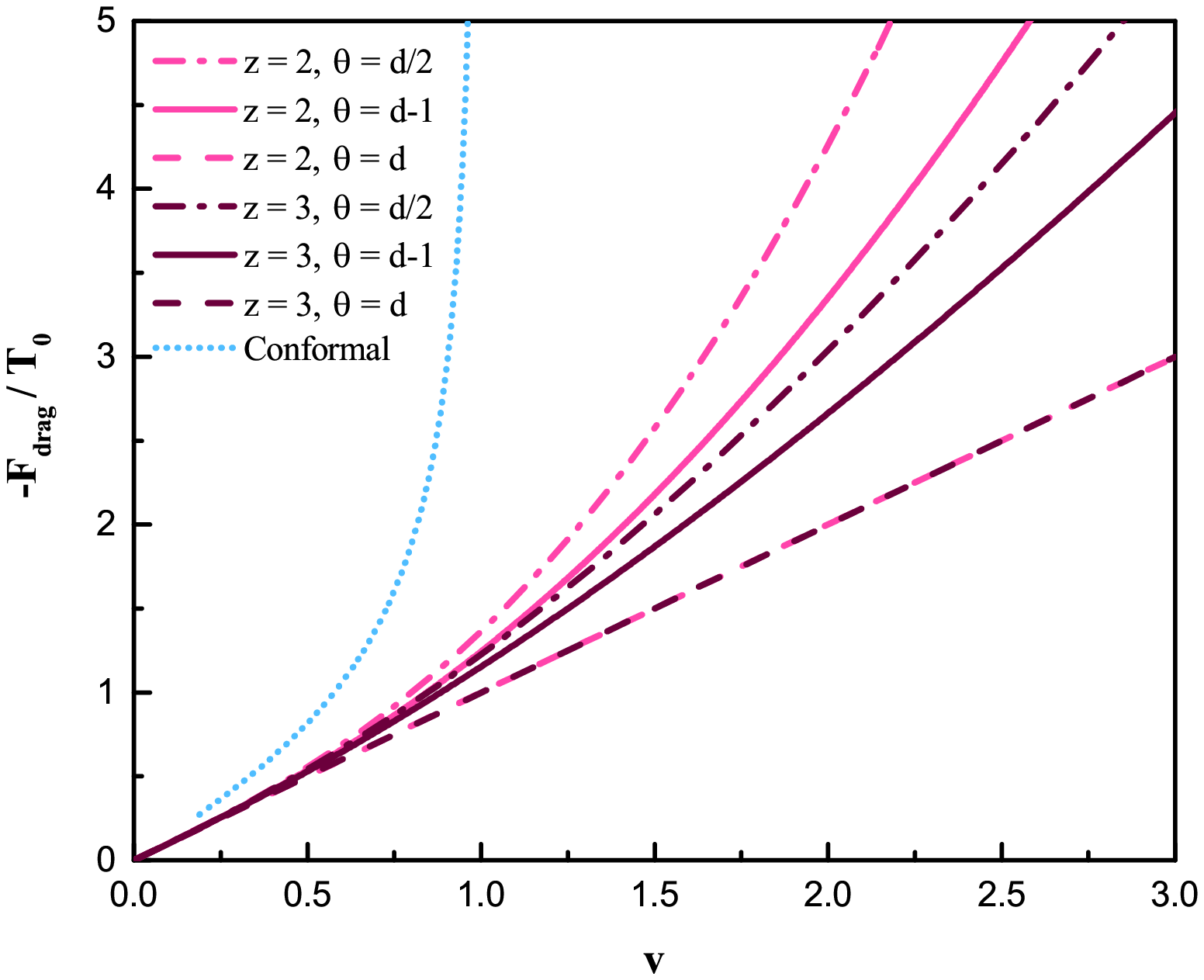}
\includegraphics[width=74 mm]{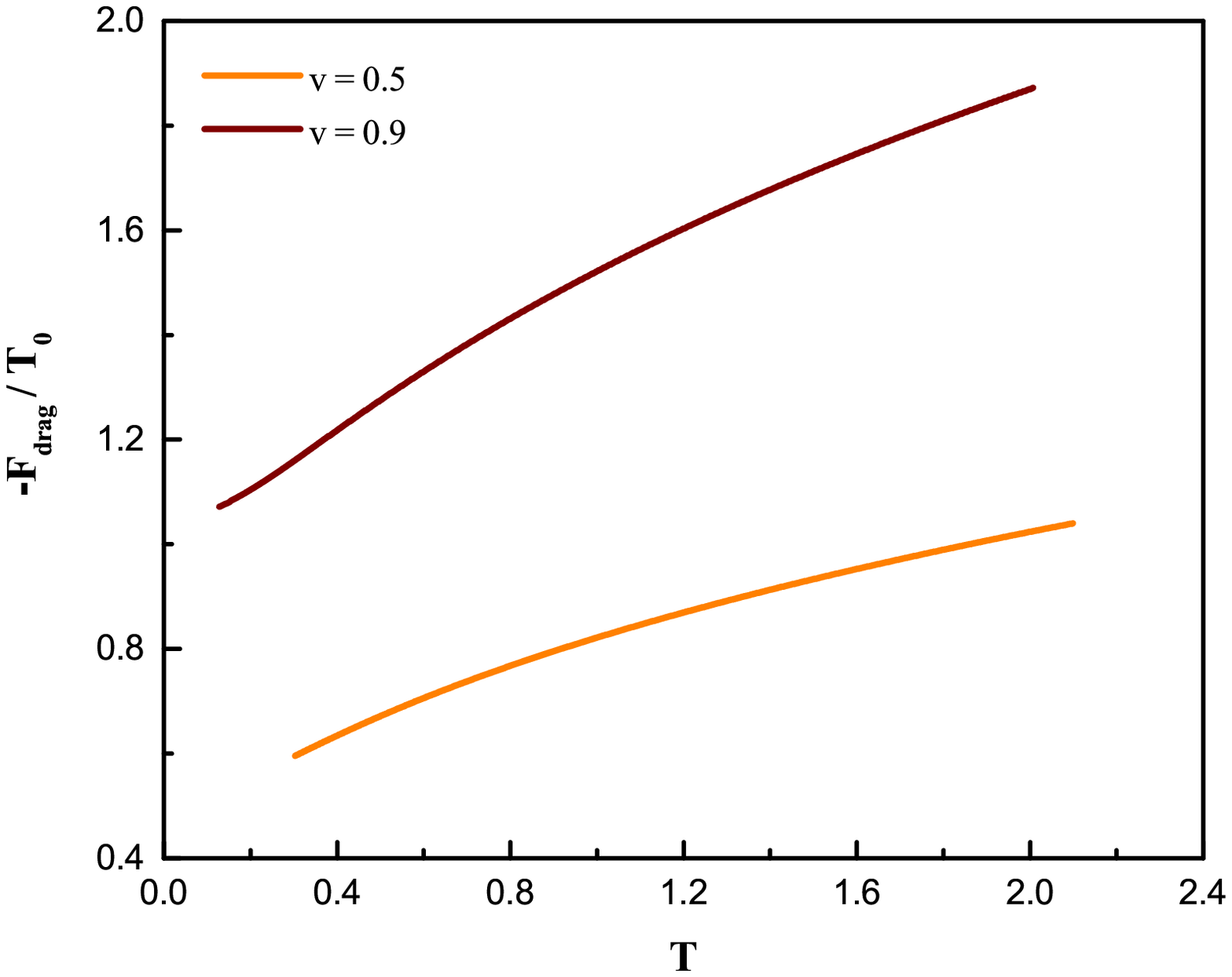}
\end{array}$
\end{center}
\caption{Left: Drag force versus velocity in different $ z $, $ \theta $ with $ m=2 $ and $ Q=1 $. Right: Drag force versus temperature at two velocities with $ z=2 $, $ \theta=d-1 $ and $ Q=1 $.}
\label{labeltem}
\end{figure}
\begin{figure}
\begin{center}$
\begin{array}{cc}
\includegraphics[width=74 mm]{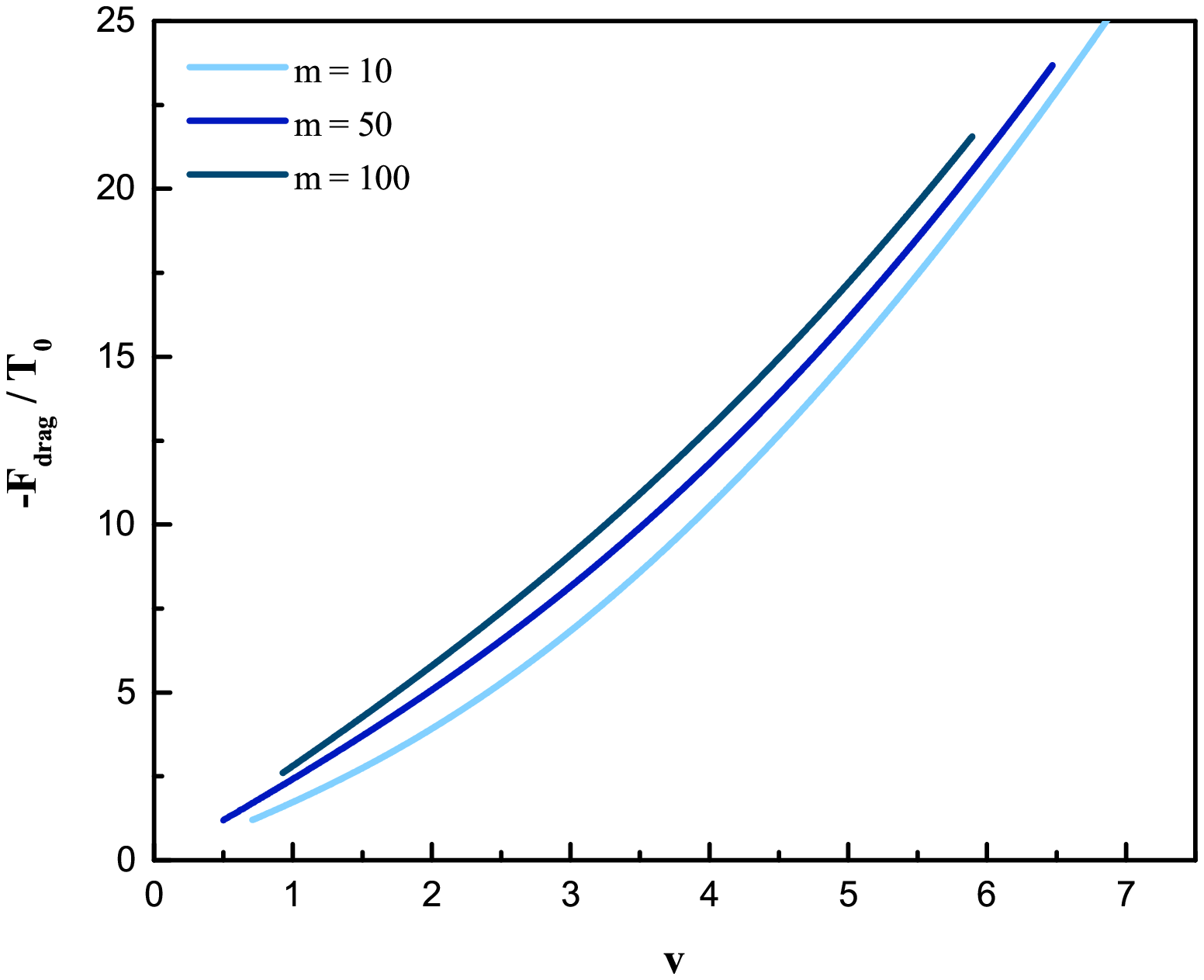}
\includegraphics[width=74 mm]{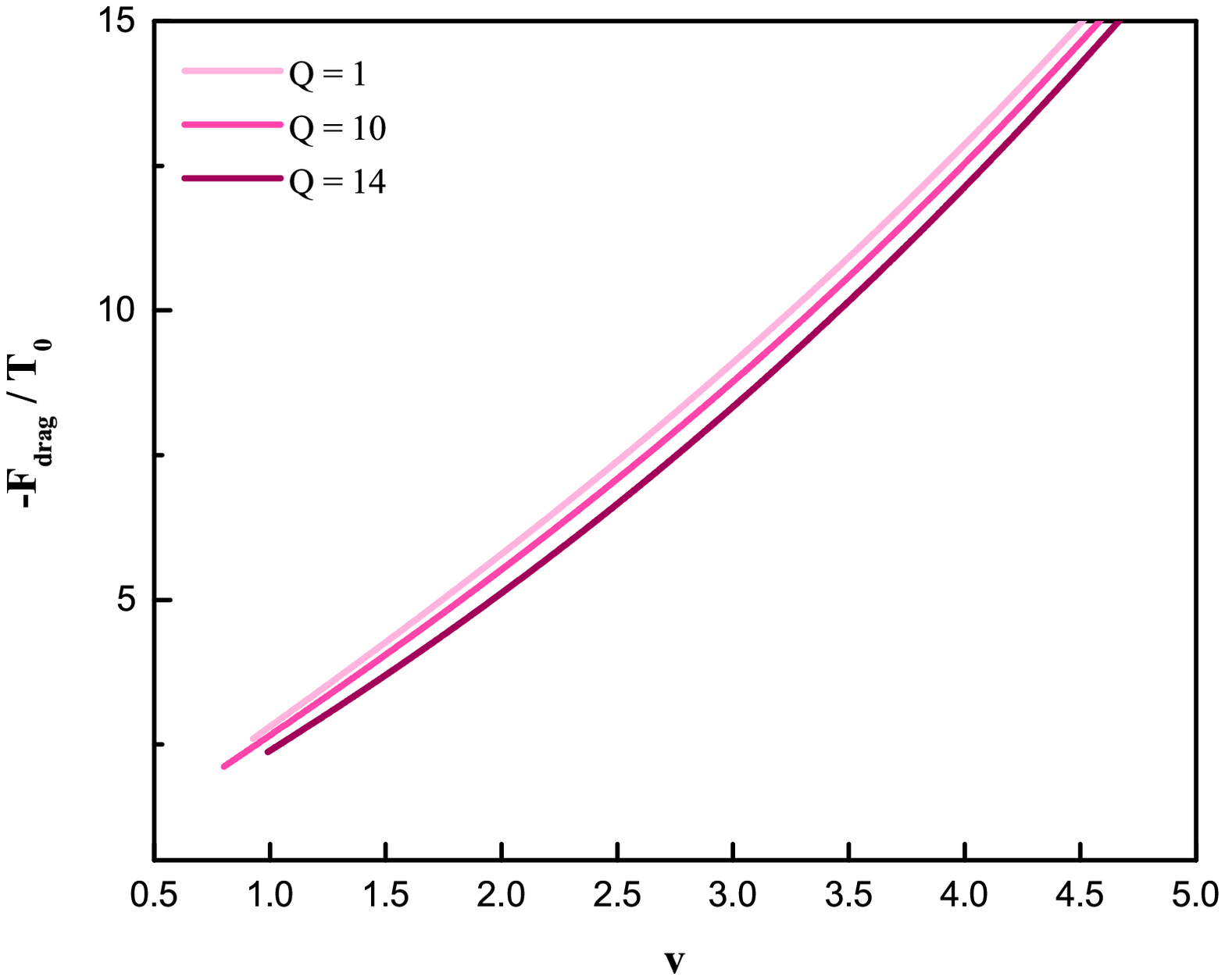}
\end{array}$
\end{center}
\caption{Left: Drag force versus velocity in different $ m $ at $ z=2 $, $ \theta =d-1 $ and $ Q=1 $. Right: Drag force versus velocity in different $ Q $ at $ z=2 $, $ \theta =d-1 $ and $ m=100 $.}
\label{label99}
\end{figure}

The heavy quark experiences the drag force that can be calculated by the momentum current along $ x^{1} $ direction. Therefore, the drag force will be following,
\begin{equation}
F_{drag}=\pi_{x}^{1}=-T_{0}vr_{c}^{2(\alpha +1)}. \label{xxx}
\end{equation}
The minus sing in $ F_{drag} $ indicates that the direction of the drag force is opposite the movement. Rewriting the drag force based on the temperature of the field theory, which was obtained from (\ref{TT}) would be complex. It is clear that in the different values $ z $ and $ \theta $ the temperature dependency of drag force is different. In Fig.[\ref{labeltem}] we plot numerically the drag force versus the velocity of the heavy quark and the temperature of the field theory. In the right plot we depict the drag force versus the velocity. As we see by increasing $ z $ and $ \theta $ the value of the drag force decreases at constant velocity. In the left plot, we draw the drag force versus the temperature in two different velocities. We see at fixed temperature, the drag force increases as the velocity increases. In Fig.[\ref{label99}], we plot the drag force versus the velocity in different values of the mass and charge. In the left plot, we see the drag force increases by increasing $ m $ at the fixed charge and velocity. In the right plot, it is clear that at fixed mass and velocity the drag force decreases as the charge increases. Moreover, in special case $ z=1 $, $ \theta=0 $ and $ Q=0 $ one can obtain the drag force for $ \mathcal{N}=4 $ case, $ F_{\mathcal{N}=4}/\pi T^{2}=-\sqrt{\lambda}v/2\sqrt{1-v^{2}} $ \cite{etl,herzog}. 

Interestingly, it was shown in \cite{heshmat} that increasing $ z $ and $ \theta $ have different effects on the jet quenching parameter. Increasing $ z $ and $ \theta $ lead to decreasing and increasing the jet quenching parameter, respectively. But we obtained that by increasing $ z $ and $ \theta $, the drag force decreases. Thus, one can concludes that the effect of the dynamical exponent $ z $ on the drag force and the jet quenching parameter are consistent, but the effect of the hyperscaling violation $ \theta $ is different. This is due to the different metric backgrounds and also the diffrenet at the considered ranges of $ z $ and $ \theta $. 

Also, one can obtain the energy loss of the quark as following, 
\begin{equation}
\dfrac{dE}{dt}=\pi_{t}^{1}=T_{0}v^{2}r_{c}^{2(\alpha +1)}.\label{d} 
\end{equation}
Substituting (\ref{c}) in (\ref{d}) one can find,
\begin{equation}
\dfrac{dE}{T_{0}dt}=f_{c}r_{c}^{2(\alpha +z)},
\end{equation}
where $ f_{c}=f(r_{c}) $. For obtaining the quark diffusion constant, we assume that the quark behaves the same as a string, with the mass $ M $ and the non-relativistic momentum $ p =Mv $. The momentum transfer rate equals to the momentum loss of the quark $ dp/dt=-\mu p $. Therefore the friction coefficient becomes,
\begin{equation}
\mu M=T_{0} r_{c}^{2(\alpha +1)},
\end{equation}
and the quark diffusion constant, $ D=\frac{T}{\mu M} $, is given by,
\begin{equation}
D=\dfrac{T}{T_{0}}r_{c}^{-2(\alpha +1)},
\end{equation}
where $ T $ is the Hawking temperature (\ref{TT}). It is crystal clear that the dependency of the diffusion coefficient to the critical radius is complicated and inevitably we should consider the special cases. Similar to \cite{dr1,zha,mada}, it is obvious that the calculated diffusion coefficient is related to the desired plasma. In the following we consider some interesting cases:
\begin{itemize}
\item $ \mathbf{\theta=d/2} $
\begin{equation}
\mu M=T_{0}r_{c},~~~~~D\! =\dfrac{T}{T_{0}}r_{c}^{-1}.
\end{equation}
\item $ \mathbf{\theta=d-1} $
\begin{equation}
\mu M=T_{0}r_{c}^{2/d},~~~~~D=\dfrac{T}{T_{0}}r_{c}^{-2/d}.
\end{equation}
\item $ \mathbf{\theta=d} $
\begin{equation}
\hspace*{-3mm}\mu M=T_{0},~~~~~ D = \dfrac{T}{T_{0}}.
\end{equation}
\end{itemize}
As we see, in both cases $ \theta=d/2 $ and $ \theta=d $, the diffusion is independent of dimension and also in the case of $ \theta=d $ is independent of $ r_{c} $. In the special case $ z=1 $, $ \theta =0 $ and $ Q=0 $, we regain the result of $ \mathcal{N}=4 $ SYM theory, i.e. $ D=2/\pi T\sqrt{\lambda} $ \cite{herzog}.

As we knew from \cite{alii} the extremal limit is given by,
\begin{equation}
r_{H}^{2(z+d-\theta -1)}=\dfrac{(z+d-\theta -2)}{z+d-\theta}Q^{2},
\end{equation}
and $ f(r) $ is given by,
\begin{equation}
f(r)=1-\dfrac{2(z+d-\theta -1)}{z+d-\theta -2}\left( \dfrac{r_{H}}{r}\right) ^{z+d-\theta}+\dfrac{z+d-\theta}{z+d-\theta -2}\left( \dfrac{r_{H}}{r}\right) ^{2(z+d-\theta -1)}.
\end{equation}
Regarding the (\ref{xxx}) and (\ref{d}), we see the drag force and the energy loss are non-zero. In other words, in this case though the system is at the zero temperature, the moving particle losses energy and after some time it will stop, because the time and direction of the motion of the string are scaled anisotropically. Indeed, the non-zero drag force will be a typical feature of the non-relativistic field theories if gravitational duals have metrics with an anisotropic scaling.

\section{Quasi-Normal Modes}
Now, we are interested in studying what occurs after long time and with the slow quark velocity. In these limits, one deals with small perturbations of a string which are known as quasi-normal modes of string world sheet, i.e. $ \dot{x} $ and $ x^{\prime} $ are small in the equation of corresponding string, so one can ignore of $ \dot{x}^{2} $ and $ x^{\prime 2} $ and find the friction coefficient term in the small velocity limit. The quasinormal modes analysis in three dimensions is studied in \cite{herzog}.\\
By considering these limits the equation (\ref{2}) changes as following,
\begin{equation}
\dfrac{\partial}{\partial r}\left( {r^{2\alpha +z+3}f(r)x'}  \right) =\dfrac{\partial}{\partial t}\left( \dfrac{r^{2\alpha -z+1}\dot{x}}{f(r)}  \right).\label{11} 
\end{equation}
We will achieve the following eigenvalue equation,
\begin{equation}
\mathcal{O}x=\mu^{2}x,\label{5} 
\end{equation}
if we assume the time-dependent solution of the form $ x(r,t)=x(r)e^{-\mu t} $. The operator $ \mathcal{O} $ will be following,
\begin{equation}
\mathcal{O}=\dfrac{f(r)}{r^{2\alpha -z+1}}\dfrac{d}{dr}r^{2\alpha +z+3}f(r)\dfrac{d}{dr}.\label{6} 
\end{equation}
By supposing $ \mu $ is small one can expand $ x(r) $ as,
\begin{equation}
x(r)=x_{0}(r)+\mu x_{1}(r)+\mu^{2} x_{2}(r)+\ldots .\label{7} 
\end{equation}
For obtaining the friction coefficient $ \mu $, we must apply Neumann boundary condition,
\begin{equation}
x^{\prime}(r_{m})=\mu x_{1}^{\prime}(r_{m})+\mu^{2}x_{2}^{\prime}(r_{m})=0.\label{8} 
\end{equation}
Substituting (\ref{6}) and (\ref{7}) in (\ref{5}), we obtain,
\begin{eqnarray}
x_{1}^{\prime}(r)&=&\dfrac{-A}{r^{2\alpha +z+3}f(r)}, \nonumber\\
x_{2}^{\prime}(r)&=&\dfrac{A}{r^{2\alpha +z+3}f(r)}\int \dfrac{r^{2\alpha -z+1}}{f(r)}dr,\label{9} 
\end{eqnarray}
where $ A $ is an integration constant. \\
Using (\ref{8}), the friction coefficient can be obtained by the following equation,
\begin{equation}
x_{1}^{\prime}(r_{m})=-\mu x_{2}^{\prime}(r_{m}),
\end{equation}
From (\ref{9}) we have,
\begin{equation}
\mu =\left( \int \dfrac{r^{2\alpha -z+1}}{f(r)}dr\right) ^{-1} \Bigg| _{r=r_{m}}.\label{mu} 
\end{equation}
For different cases we find,
\begin{itemize}
\item $ \mathbf{\theta=d/2} $
\begin{equation}
\mu =\left( \int \dfrac{dr}{r^{z}-\frac{m}{r^{d/2}}+\frac{Q^{2}}{r^{z+d-2}}}\right)^{-1}.
\end{equation}
\item $ \mathbf{\theta=d-1} $
\begin{equation}
\mu =\left( \int \dfrac{dr}{r^{z+1-2/d}-\frac{m}{r^{2/d}}+\frac{Q^{2}}{r^{z-1+2/d}}}\right)^{-1}.
\end{equation}
\item $ \mathbf{\theta=d} $
\begin{equation}
\mu =\left( \int \dfrac{dr}{r^{z+1}-rm+\frac{Q^{2}}{r^{z-3}}}\right)^{-1}.
\end{equation}
In this case by considering $ z=2 $ we obtain,
\begin{equation}
\mu =\dfrac{2(m-Q^{2})}{\ln\left( \dfrac{r_{H}^{2}(r_{m}^{2}-m+Q^{2})}{r_{m}^{2}(r_{H}^{2}-m+Q^{2})}\right) },~~~~~~D=\dfrac{T}{2M(m-Q^{2})}\ln\left( \dfrac{r_{H}^{2}(r_{m}^{2}-m+Q^{2})}{r_{m}^{2}(r_{H}^{2}-m+Q^{2})} \right).
\end{equation}
\end{itemize}
As we know from \cite{herzog}, the equation of motion a particle that moves with momentum $ p $ in a viscous medium and is influenced by a driving force $ f $, is given by $ \dot{p}=-\mu p+f $. In our case that $ f $ equals to zero, we have $ \dot{p}=-\mu p $. Using the latter equation and (\ref{mu}), one can obtain the drag force.

In order to obtain the total momentum of string, we calculate $ \pi_{x}^{0} $ from (\ref{11}) and put in the equation (\ref{4}). Then, we integrate from the boundary to an IR cutoff $ r_{min} >r_{H} $ and apply Neumann boundary condition. So we have,
\begin{equation}
p= \dfrac{T_{0}}{\mu}\left[ r_{min}^{2\alpha +z+3}f(r_{min})x^{\prime}(r_{min})\right] .\label{12} 
\end{equation}
Considering $ z=2 $ we find,
\begin{eqnarray}
p&=&\dfrac{T_{0}}{\mu} r_{min}^{2} \left[ r_{min}^{2}-\dfrac{m}{r_{min}^{d/2}}+\dfrac{Q^{2}}{r_{min}^{d}} \right] x^{\prime}(r_{min}),~~~~ (For~ \theta =d/2),\nonumber\\
p&=&\dfrac{T_{0}}{\mu} r_{min}^{2/d} \left[ r_{min}^{3}-m+\dfrac{Q^{2}}{r_{min}} \right] x^{\prime}(r_{min}),~~~~~~~~ (For~ \theta =d-1),\nonumber\\
p&=&\dfrac{T_{0}}{\mu} r_{min} \left[ r_{min}^{2}-(m-Q^{2})\right] x^{\prime}(r_{min}),~~~~~~~~~~ (For~ \theta =d).
\end{eqnarray}
Now, we want to obtain the total energy. For this purpose, we must expand $ (-g)^{-1/2} $ to second order of velocities, then by using (\ref{11}) we obtain the following equation,
\begin{equation}
\pi_{t}^{0}=-T_{0}\left( r^{2\alpha +z-1}+\frac{1}{2}\left( r^{2\alpha +z+3}f(r)x x^{\prime}\right) ^{\prime}\right).
\end{equation}
Finally, the total energy becomes,
\begin{equation}
E=T_{0}\left( \frac{1}{2\alpha +z}\left( r_{m}^{2\alpha +z}-r_{min}^{2\alpha +z}\right) -\dfrac{1}{2}r_{min}^{2\alpha +z+3}f(r_{min})x(r_{min})x^{\prime}(r_{min})\right),\label{13} 
\end{equation}
where we also used Neumann boundary condition. Again we apply different hyperscaling violation parameters which are given by:
\begin{itemize}
\item $ \mathbf{\theta=d/2} $ \\
If $ z=2 $ then one can obtain the following equation,
\begin{equation}
E=T_{0}\left( r_{m}-r_{min}-\dfrac{r_{min}^{2}}{2}\left( r_{min}^{2}-\dfrac{m}{r_{min}^{d/2}}+\dfrac{Q^{2}}{r_{min}^{d}}\right) x(r_{min}) x^{\prime}(r_{min})\right).
\end{equation}
\item $ \mathbf{\theta=d-1} $ \\
Considering $ z=2 $ one have,
\begin{equation}
E=\frac{T_{0}}{2}\left( d\left( r_{m}^{2/d}-r_{min}^{2/d}\right) -r_{min} ^{2/d}\left( r_{min}^{3}-m+\frac{Q^{2}}{r_{min}}\right) x(r_{min}) x^{\prime}(r_{min})\right).
\end{equation}
\item $ \mathbf{\theta=d} $ \\
In case that $ z=2 $ and $ \theta=d $, $ E $ becomes infinty. Thus, by considering $ z=3 $, then one find,
\begin{equation}
E=T_{0}\left( r_{m}-r_{min}-\dfrac{1}{2}\left( r_{min}^{4}-mr_{min}+Q^{2}\right) x(r_{min}) x^{\prime}(r_{min})\right).
\end{equation}
\end{itemize}

Substituting (\ref{12}) in (\ref{13}) and using $ p=M\dot{x}=-\mu Mx $, where $ M $ is the kinetic mass of the quark, we achieve a simple relation between the energy and momentum, $ E=\frac{T_{0}}{2\alpha +z}\left( r_{m}^{2\alpha +z}-r_{min}^{2\alpha +z}\right)+\frac{p^{2}}{2M} $, that in the special case $ \theta=d $, should $ z>2 $ . We consider ﬁrst term of right hand side as $ M_{rest} $ of the quark.  

\section{Adding Constant Electromagnetic Field to the Background}
In the previous sections, we calculated the drag force in the background with hyperscaling violation. We know the velocity of a particle, that firstly was constant, decreases by time passing due to loss of energy. Therefore its motion turns into the small ﬂuctuations. Now, we want to study the effect of non-zero electromagnetic field on the drag force following the method proposed in \cite{field}. This constant B-field that is $ B=Edt\wedge dx_{1}+\mathcal{H}dx_{1}\wedge dx_{2} $, couples to line element (\ref{1}). $ E $ and $ \mathcal{H} $ are the NS-NS antisymmetric electric and magnetic fields along the $ x_{1} $ and $ x_{2} $ directions and other components of B-field are zero. Due to that only the field strength is involved in equation of motion, this ansatz is a good solution to supergravity and the minimal setup to study the B-field correction. The dual D-string with a monopole may be shown by these solutions, $ x_{1}(r,t)=x_{1}(r)+v_{1}t $, $ x_{2}(r,t)=x_{2}(r)+v_{2}t $ and $ x_{3}(r,t)=0 $. Therefore the corresponding Lagrangian density will be as,
\begin{equation}
\mathcal{L}=\left[ r^{4\alpha} \left(  r^{2z-2}-\frac{|\vec{v}|^{\!\! ~2}}{f(r)}+r^{2z+2}f(r) |\vec{x}^{\prime}|^{2}-r^{4}|\vec{v} \times \vec{x}^{\prime}|^{2}
\right)-\left( Ex_{1}^{\prime}+\mathcal{H}\vec{v}\times \vec{x}^{\prime}\right)^{2} \right]^{1/2}.
\end{equation}

In the first case, we assume $ \mathcal{H}=0 $, it means that only electric ﬁeld exists and the monopole moves with constant speed along the $ x_{1} $ direction. Therefore, our solutions are restricted to $ x_{1}(r,t)=x_{1}(r)+vt $ and $ x_{2}(r,t)=x_{3}(r,t)=0 $. Using of $ \pi_{x_{1}}^{1}=\frac{\partial \mathcal{L}}{\partial x^{\prime}_{1}} $ and the reality condition for $ x^{\prime}_{1} $, we can obtain the $ x_{1} $-component of the momentum density and the drag force which are as following,
\begin{eqnarray}
\pi_{x_{1}}^{1}&=&\left( v^{2}r_{c}^{4(\alpha +1)}-E^{2}\right) ^{\frac{1}{2}},\nonumber\\
\dfrac{dp}{dt}&=&-T_{0}\left( v^{2}r_{c}^{4(\alpha +1)}-E^{2}\right) ^{\frac{1}{2}},
\end{eqnarray}
where $ r_{c} $ is the root of equation (\ref{c}). As we see, the drag force is decreased by the electric ﬁeld $ E $, this is in agreement with the reported results in \cite{field}. If $ E=0 $, we regain the equation (\ref{d}) and if $ E= vr_{c}^{2(\alpha +1)} $, then $ \frac{dp}{dt}=0 $ and D-string feels no drag force. This value is the amount that needed to the velocity of particle remain constant.
\begin{eqnarray}
\dfrac{dp}{dt}&=&-T_{0}\left(v^{2}r_{c}^{2}-E^{2}\right)^{1/2},~~~~~~~~~(For~ \theta=d/2),\nonumber\\
\dfrac{dp}{dt}&=&-T_{0}\left(v^{2}r_{c}^{4/d}-E^{2}\right)^{1/2},~~~~~~~\!(For~\theta=d-1),\nonumber\\
\dfrac{dp}{dt}&=&-T_{0} \left(v^{2}-E^{2}\right)^{1/2},~~~~~~~~~~~~(For~\theta=d).
\end{eqnarray}

In the second case, we consider the constant magnetic field $ \mathcal{H} $ and choose the solutions $ x_{1}(r,t)=x_{1}(r)+vt $, $ x_{2}(r,t)=x_{2}(r) $ and $ x_{3}(r,t)=0 $. For this case, we obtain,
\begin{eqnarray}
x_{1}^{\prime}&=&\sqrt{\dfrac{\pi_{x_{1}}^{2}\left( r^{\beta}f\right) ^{-1}\left( r^{\beta -4} -r^{4\alpha}v^{2}f^{-1}\right) \left( r^{\beta}f- v^{2}( r^{4(\alpha +1)}+\mathcal{H}^{2}) \right) }
{ \left( \pi_{x_{1}}^{2}-r^{\beta}f\right) \left( \pi_{x_{2}}^{2}-r^{\beta}f+v^{2} (r^{4(\alpha +1)}+\mathcal{H}^{2})\right) 
-\pi_{x_{1}}^{2}\pi_{x_{2}}^{2} }},\nonumber\\
x_{2}^{\prime}&=&\sqrt{\dfrac{\pi_{x_{2}}^{2}\left( r^{\beta}f \right) \left( r^{\beta -4}-r^{4\alpha}v^{2}f^{-1}\right) \left( r^{\beta}f-v^{2} (r^{4(\alpha +1)}+\mathcal{H}^{2})\right)^{-1}}
{ \left( \pi_{x_{1}}^{2}-r^{\beta}f\right) \left( \pi_{x_{2}}^{2}-r^{\beta}f+v^{2} (r^{4(\alpha +1)}+\mathcal{H}^{2})\right) 
-\pi_{x_{1}}^{2}\pi_{x_{2}}^{2}}},
\end{eqnarray}
where $ \pi_{x_{i}}\equiv \pi_{x_{i}}^{1} $. By applying the reality condition one can calculate two momentum densities as $ \pi_{x_{1}} $ and $ \pi_{x_{2}} $,
\begin{equation}
\pi_{x_{1}}^{1}=vr_{c}^{2(\alpha +1)},~~~~~\pi_{x_{2}}^{2}=0,
\end{equation}
and also the drag forces will be as,
\begin{equation}
\dfrac{dp_{1}}{dt}=-T_{0}v r_{c}^{2(\alpha +1)},~~~~~ \dfrac{dp_{2}}{dt}=0.
\end{equation}
As we see, in $ x_{2} $ direction the drag force is zero, it means that $ \mathcal{H} $ dose not affect on the motion along $ x_{1} $ direction. In fact, since there is no the movement along $ x_{2} $ direction, the corresponding drag force will be vanished. Also, because the string cannot move along $ x_{3} $, the magnetic field has not effect on the motion along the direction related to the Lorentz force. Therefore, just the appropriate electric field has effect on the motion of the string and keeps it at conatant speed $ v $ that this is in agreement with results of \cite{gubser,herzog}.

In the last case, we consider $ \vec{v}\perp E $. In this case, we achieve two solutions. The first solution is $ \pi_{1}=0 $ therefore we have no the drag force along the moving direction. The second solution is $ \pi_{2}=0 $, it implies that the electric ﬁeld $ E $ has no eﬀect on the drag force.

\section{The Energy loss of rotating string}
In this section, we want to study the energy loss of a rotating quark. This quark that rotates along a circle of radius $ L $ with a constant angular frequency $ \omega $, corresponds to a open string hanging of the boundary which has the motion of spiral in the bulk. For calculating the energy loss of this string, one should find the world-sheet of the spiraling string and then can calculate the energy ﬂowing down the string \cite{rot,rott}.\\
Firstly, we change the coordinates of the metric background (\ref{1}) as,
\begin{equation}
dx_{1}^{2}+dx_{2}^{2}=d\rho^{2}+\rho^{2}d\phi^{2}, ~~~~~~ where~~~~ (x_{1},x_{2})\rightarrow (\rho,\phi).
\end{equation}
Then, we can rewrite the metric (\ref{1}) as,
\begin{equation}
ds^{2}=r^{2\alpha}\left( -r^{2z}f(r)dt^{2}+\dfrac{dr^{2}}{r^{2}f(r)}+r^{2}\left( d\rho^{2}+\rho^{2}d\phi^{2}\right) +r^{2}d{\bf{x}}^{2}\right).
\end{equation}
We parameterize the world-sheet of the spiraling string $ X^{\mu}(\tau,\sigma) $ as,
\begin{equation}
X^{\mu}(\tau,\sigma)=\left(  t=\tau, \rho=\rho(r), \phi= \omega\tau +\varphi(r), x_{3}=0, r=\sigma\right),
\end{equation}
where $ \rho(r) $ and $ \varphi(r) $ denote the radial and angular proﬁles of the rotating string respectively. They should satisfy the following boundary
conditions,
\begin{equation}
\rho(\infty)=L,~~~~ \varphi(\infty)=0. \label{mr} 
\end{equation}
As we know the Nambu-Goto action which describes dynamics of such string, will be as,
\begin{equation}
S=-T_{0}\int dtdr~r^{2\alpha}\sqrt{\left( r^{2z}f-\omega^{2}r^{2}\rho^{2} \right) \left( \frac{1}{r^{2}f}+r^{2}\rho^{\prime 2}\right) +r^{2(z+1)}f\rho^{2}\varphi^{\prime 2}},
\end{equation}
where $ \prime $ is derivative with respect to the radial coordinate $ r $. We then obtain the equations of motion as following,
\begin{equation}
\varphi^{\prime 2}=\dfrac{\Pi_{\varphi}^{2}\left( r^{2z}f-\omega^{2}r^{2}\rho^{2} \right) \left( (r^{2}f)^{-1}+r^{2}\rho^{\prime 2}\right)}{r^{2(z+1)}f\rho^{2}\left( r^{\beta}f\rho^{2}-\Pi_{\varphi}^{2} \right) }, \label{F} 
\end{equation}
\begin{eqnarray}
\rho^{\prime\prime}\!\! &&+\left[  \dfrac{\left( r^{2z}f-\omega^{2}r^{2}\rho^{2} \right) \left( (r^{2}f)^{-1}+r^{2}\rho^{\prime 2}\right)}{\left( r^{2z}f-\omega^{2}r^{2}\rho^{2} \right)^{2}}\right]  \Bigg[ \rho \omega^{2}-\dfrac{\Pi_{\varphi}^{2}f\left(r^{2z}f -\omega^{2}r^{2}\rho^{2} \right) \left( (r^{2}f)^{-1}+r^{2}\rho^{\prime 2}\right)}{\rho\left(r^{\beta}f\rho^{2} -\Pi_{\varphi}^{2}\right) } \nonumber\\&&+ f\rho^{\prime}\left( r^{2(z+1)}f \left( \dfrac{2\alpha +z+1}{r}+\dfrac{f^{\prime}}{f}\right) 
-\omega^{2}r^{4}\rho^{2}\left( \dfrac{2\alpha+4}{r}+\dfrac{\rho^{\prime}}{\rho}\right) \right) \nonumber\\ &&+\left( \dfrac{2\alpha+z+1}{r}+\dfrac{\rho^{\prime}}{\rho}+\dfrac{f^{\prime}}{2f} \right) 
\left( \dfrac{\Pi_{\varphi}^{2}r^{2}f\rho^{\prime}( r^{2z}f-\omega^{2}r^{2}\rho^{2})}{r^{\beta}f\rho^{2}-\Pi_{\varphi}^{2}} \right) \Bigg] \nonumber\\ &&-\dfrac{r^{2}f\rho^{\prime}}{r^{2z}f-\omega^{2}r^{2}\rho^{2}} \Bigg[ \dfrac{z-1}{r}r^{2(z-1)}+r^{2(z+1)}f\rho^{\prime 2}\left( \dfrac{z+1}{r}+\dfrac{f^{\prime}}{2f}\right) \nonumber\\&&-\omega^{2}\rho^{2}\left( \dfrac{1}{f}\left( \dfrac{\rho^{\prime}}{\rho}-\dfrac{f^{\prime}}{2f}\right)+r^{4}\rho^{\prime2} \left( \dfrac{2}{r}+\dfrac{\rho^{\prime}}{\rho}\right)  \right) \Bigg]=0. \label{P} 
\end{eqnarray}
Here, we note that the metric (\ref{1}) is not dependent on the angular variable $ \varphi $ explicitly. So the Nambu-Goto action is independent of $ \varphi $, therefore the conjugate angular momentum, $ \Pi_{\varphi}=\frac{\partial \mathcal{L}}{\partial\varphi^{\prime}} $, is the constant of motion. Using the latter equation, we could obtain the equation of motion $ \varphi $, as equation (\ref{F}). This equation is ﬁrst order, thus the boundary condition (\ref{mr}) is enough to determine the full solution. On the other hand, the metric (\ref{1}) is dependent on $ \rho $ and hence its equation of motion will be of second order. By using (\ref{F}), we have obtained the equation of motion $ \rho(r) $ as (\ref{P}). In general, finding $ \rho(r) $ is not easy and one must solve the diﬀerential equation (\ref{P}) numerically. The simplest case study about the rotating string is $ \mathcal{N}=4 $ SYM plasma \cite{rott}.

\begin{figure}
\begin{center}$
\begin{array}{cc}
\includegraphics[width=73 mm]{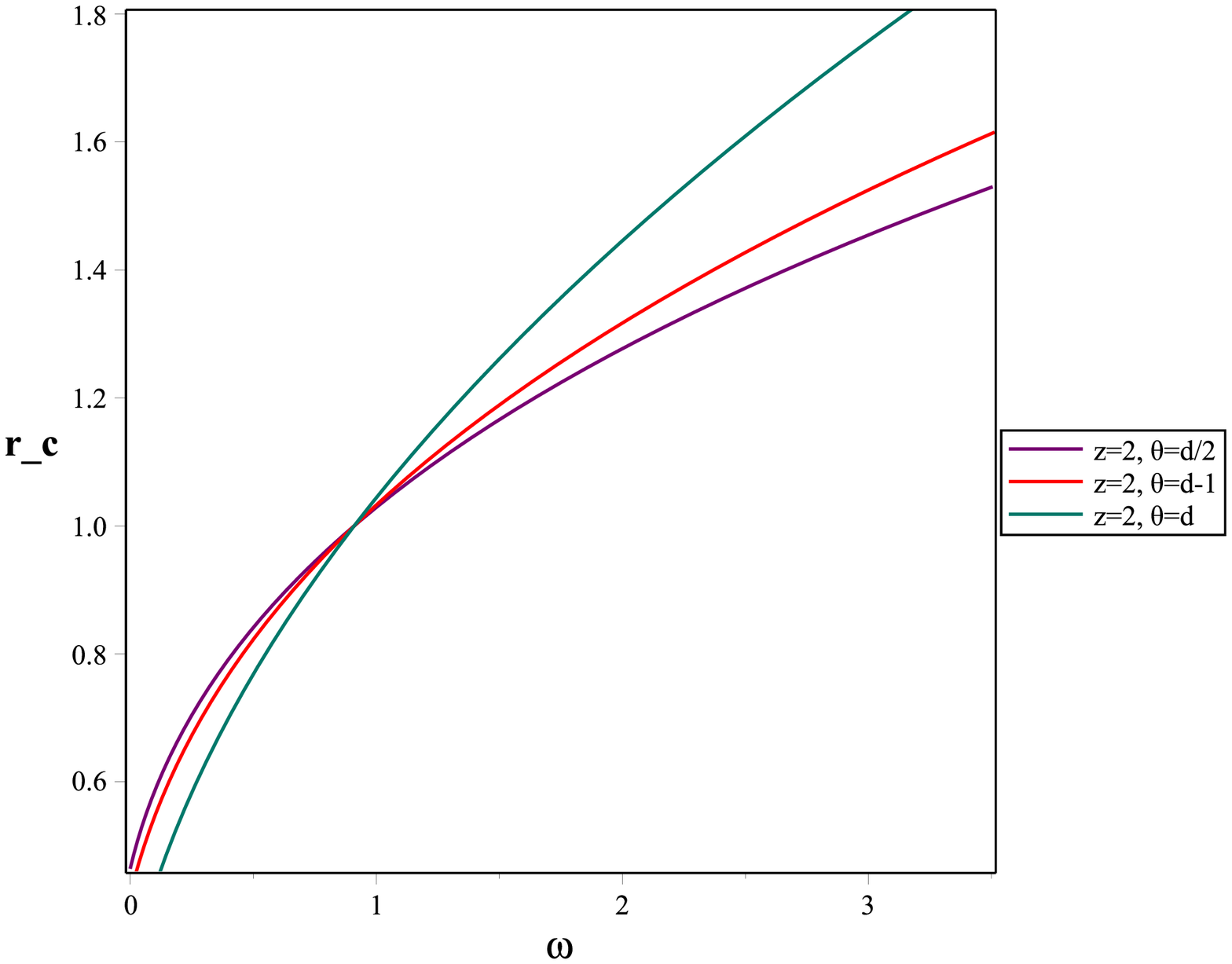}
\includegraphics[width=73 mm]{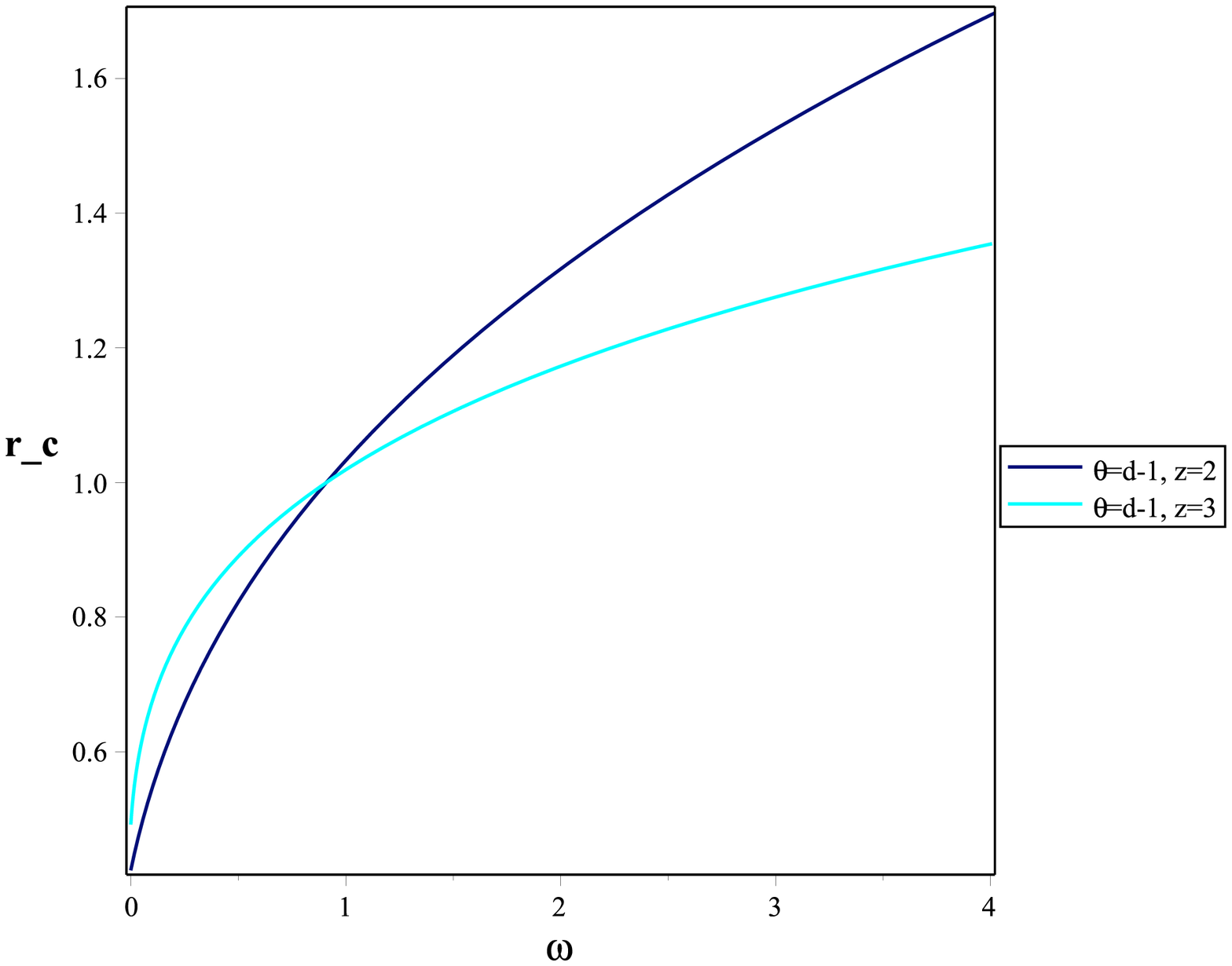}
\end{array}$
\end{center}
\caption{Worldsheet horizon $ r_{c} $ versus $ \omega $ at $ m=0.1, Q=0.1$ and $ \Pi=1 $ in different $ z $ (left plot), in different $ \theta $ (right plot).}
\label{lab4}
\end{figure}

As we saw, the equation (\ref{P}) is a second order diﬀerential equation, so it requires two boundary conditions to have solution uniquely. One of these conditions is given by the equation (\ref{mr}). The other one can be obtained by imposing the reality condition on $ \varphi^{\prime} $. Since the LHS of (\ref{F}) is always positive deﬁnite, the numerator and denominator in the RHS must change signs at the same point in order to be positive deﬁnite. Also, $ \rho $ is singular when $ r^{2z}f-\omega^{2}r^{2}\rho^{2}=0 $ or $ r^{\beta}f\rho^{2}-\Pi_{\varphi}^{2}=0 $. We denote this point by $ r_{c} $, and we have,
\begin{equation}
\rho_{c}=\sqrt{\dfrac{\Pi_{\varphi}}{\omega r^{2(\alpha+1)}_{c}}}=\sqrt{\dfrac{r^{2(z-1)}_{c}f_{c}}{\omega^{2}}},~~~~~~  r^{2(\alpha+z)}_{c}f_{c}=\omega\Pi_{\varphi},\label{kl} 
\end{equation}
where $ f_{c} $ is the function $ f(r) $ in $ r_{c} $ and $ \rho_{c} $ is the radius of spiraling string at this value. In Fig. [\ref{lab4}], we have plotted $ r_{c} $ in terms of changing angular frequency $ \omega $ and dynamical exponents $ z $ and $ \theta $. As we see the behaviors of $ z $ and $ \theta $ before and after $ \omega=1 $ are completely different. In the left plot, at $ \omega <1 $ it is obvious by increasing $ \theta $ the critical radius $ r_{c} $ decreases and increases in the regime $ \omega >1 $. In the right plot, at $ \omega <1 $ we see by increasing $ z $ in the special $ \theta=d-1 $, $ r_{c} $ increases and decreases in the regime $ \omega >1 $.

The most important point here is a horizon of the string. When it appears the string coils in the bulk and $ r_{c} $ places in special point. Also there is no any black hole in the bulk but there is the horizon on the string world-sheet.

In order to finding the energy loss rate, we use the world-sheet momentum, and therefore we find,
\begin{equation}
\dfrac{dE}{dt}=\dfrac{\omega r^{2(\alpha +z+1)}f\rho^{2}\varphi^{\prime}}{2\pi\alpha^{\prime}\sqrt{\left( r^{2z}f-\omega^{2}r^{2}\rho^{2} \right) \left( \frac{1}{r^{2}f}+r^{2}\rho^{\prime 2}\right) +r^{2(z+1)}f\rho^{2}\varphi^{\prime 2}}}=\dfrac{\omega\Pi_{\varphi}}{2\pi\alpha^{\prime}}.\label{kj} 
\end{equation}
Substituting (\ref{kl}) in (\ref{kj}), we obtain,
\begin{equation}
\dfrac{dE}{dt}=\dfrac{r^{2(\alpha+z)}_{c}f_{c}}{2\pi\alpha^{\prime}}.
\end{equation}
\begin{figure}
\begin{center}$
\begin{array}{cc}
\includegraphics[width=74 mm]{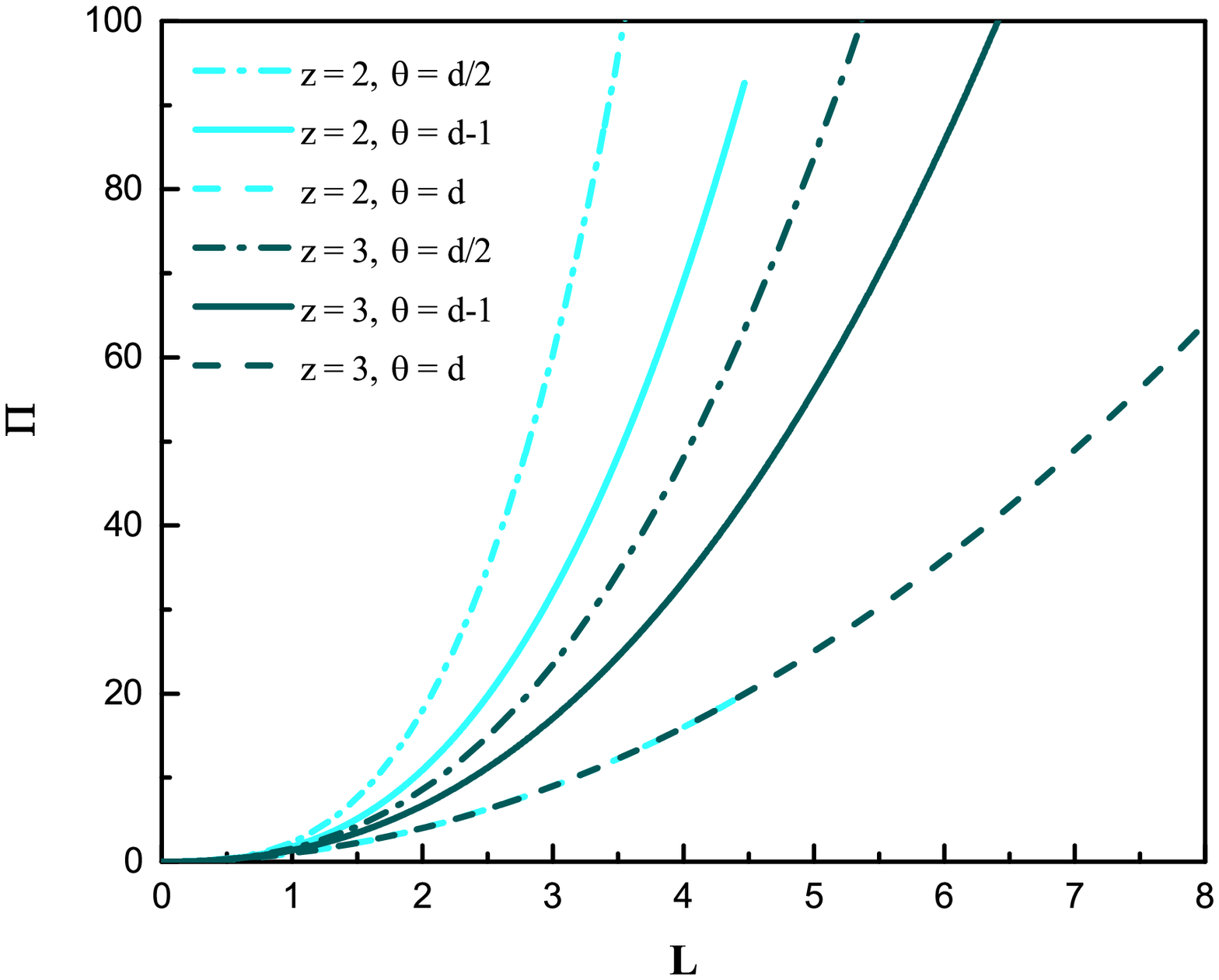}
\includegraphics[width=74 mm]{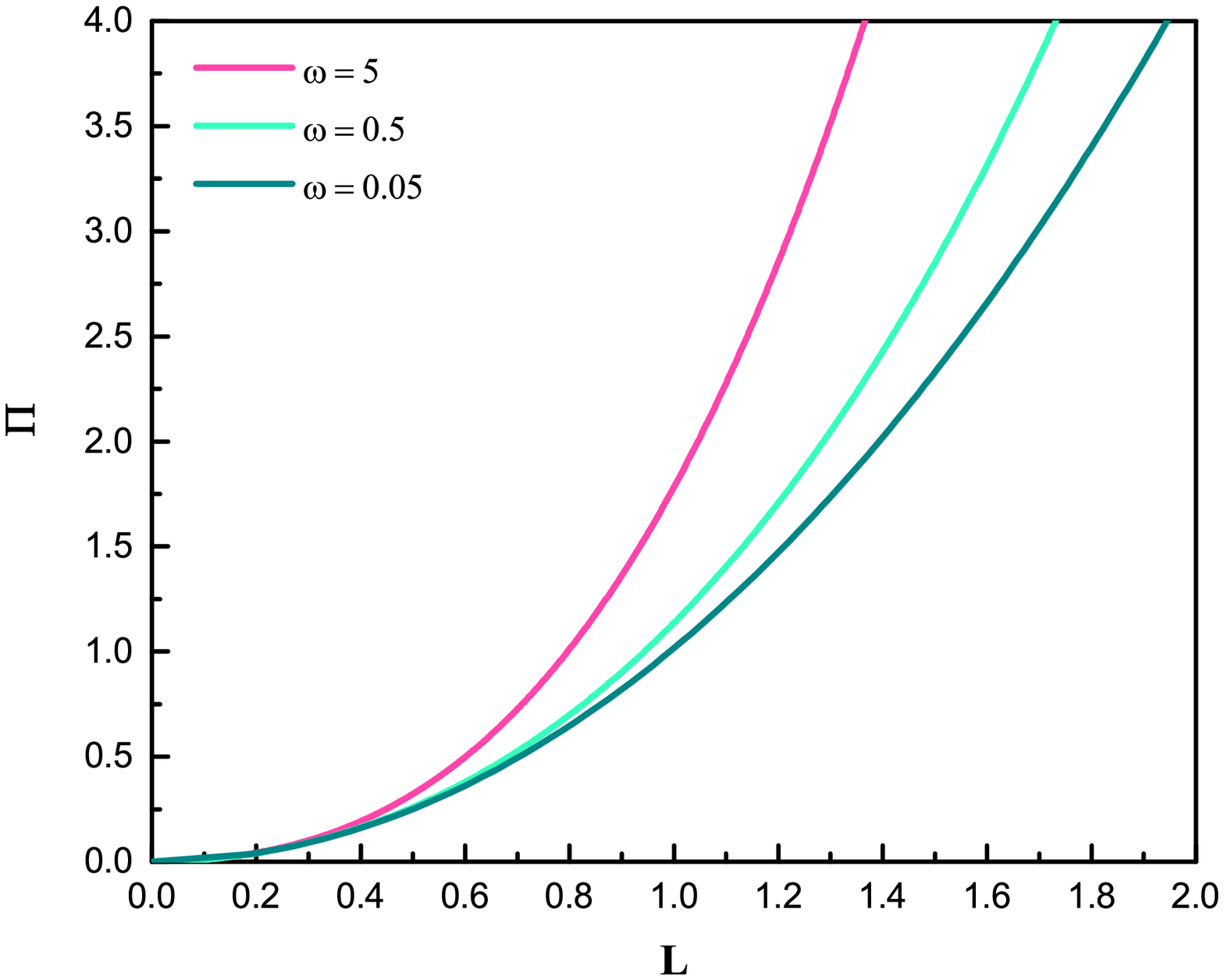}
\end{array}$
\end{center}
\caption{Left: Energy loss versus rotation radius of quark in different $ z $ and $ \theta $ at $ \omega=5 $, $ m=2 $ and $ Q=1 $. Right: Energy loss versus rotation radius in different $ \omega $ at $ z=2 $, $ \theta=d-1 $, $ m=2 $ and $ Q=1 $.}
\label{ww}
\end{figure}
This equation is the same as the case in which the quark moved with constant velocity. It is interesting to compute the energy loss in terms of the field theory parameters $ \omega $ and $ L $. Numerically, we plot the energy loss of the rotating particle versus the radius of the rotation in Fig. [\ref{ww}]. In the left plot we see as $ z $ and $ \theta $ increases at the constant $ \omega=5 $, the energy loss decreases. In the right plot, we consider $ z=2 $ and $ \theta=d-1 $ and plot $ \Pi $ versus $ L $ for different values of $ \omega $. It is obvious by increasing $ \omega $ the energy loss increases.

\section{Discussion}
In this paper, we considered a moving quark in a non-relativistic background. This geometry can be obtained from an Einstein-Maxwell-Dilaton theory so that the background consists both abelian gauge fields and scalar fields. Therefore it has two dynamical and hyperscaling violation parameters. 

Using AdS/CFT correspondence we calculated the drag force and the energy loss of the moving quark in various values $ z $ and $ \theta $ and studied effects of $ m $ and $ Q $. We plotted the drag force versus the velocity and the temperature. We saw the drag force increases by increasing the velocity at the constant $ m $, $ Q $, $ z $ and $ \theta $ and decreases by increasing $ z $ and $ \theta $ at the constant $ m $, $ Q $ and $ v $. The drag force increases by increasing the mass and decreases by increasing the charge at the special case $ z=2 $ and $ \theta=d-1 $. In the drag force versus the temperature plot, we saw the drag force increases by increasing velocity at fixed temperature. Also we investigated the quasi normal modes and added a electromagnetic field on the brane. The constant electric field decreases the corresponding drag force and the constant magnetic field has no effect which these results are in accordance with \cite{gubser,herzog,field}. In the last section, we considered the rotating quark and calculated its the energy loss. We saw that the energy loss of the rotating and moving string is similar. We saw the drag force increases by increasing $ \omega $ and decreases by increasing $ z $ and $ \theta $. We observed for the moving or rotating open string the energy loss is non-zero even at zero temperature, because the event horizon appears on worldsheet of this string.

\end{document}